\documentclass[showpacs,showkeys,preprintnumbers,amsmath,amssymb]{revtex4}

\usepackage{graphicx}

\usepackage{dcolumn}
\usepackage{bm}

\providecommand{\beqa}{\begin{eqnarray}}
 \providecommand{\bf}{\mathbf}
 \providecommand{\rm}{\mathrm}
\providecommand{\eeqa}{\end{eqnarray}}

\def\Z2{{\mathbf{Z}_2}}
 
 \def\tg{{\tilde{\Gamma}}}
 \def\vepsn{{\varepsilon}}

\begin{document}


\title{Observing the Structure of the Landscape with the CMB Experiments}

\author{Amjad Ashoorioon}
\email{amjad.ashoorioon@fysast.uu.se}

\affiliation{Institutionen f\"{o}r fysik och astronomi
Uppsala Universitet, Box 803, SE-751 08 Uppsala, Sweden}


\date{\today}

\begin{abstract}

Assuming that inflation happened through a series of tunneling in the string theory landscape, it is argued that one can determine the structure of vacua using precise measurements of the scalar spectral index and tensor perturbations at large scales. It is shown that for a vacuum structure where the energy gap between the minima is constant, {\it i.e.} $\epsilon_i=i m_f^4$, one obtains the scalar spectral index, $n_s$, to be $\simeq 0.9687$, for the modes that exit the horizon $60$ e-folds before the end of inflation. Alternatively, for a vacuum structure in which the energy gap increases linearly with the vacuum index, \textit {i.e.} $\epsilon_i=\frac{i^2}{2} m_f^4$,  $n_s$ turns out to be $\simeq 0.9614$.  Both these two models are motivated within the string theory landscape using flux-compactification and their predictions for scalar spectral index are compatible with WMAP results. For both these two models, the results for the scalar spectral index turn out to be independent of $m_f$. Nonetheless, assuming that inflation started at Planckian energies and that there had been successful thermalization at each step, one can constrain $m_f<2.6069\times 10^{-5} m_P$ and $m_f<6.5396\times 10^{-7} m_P$ in these two models, respectively. Violation of the single-field consistency relation between the tensor and scalar spectra is another prediction of chain inflation models. This corresponds to having a smaller tensor/scalar ratio at large scales in comparison with the slow-roll counterparts. Similar to slow-roll inflation, it is argued that one can reconstruct the vacuum structure using the CMB experiments.

\end{abstract}


\maketitle
\section{Introduction}
In the original picture of chain inflation, as proposed by \cite{Freese:2004vs}, the universe tunnels rapidly
through a series of first order phase transitions.  During the time spent in any
one of these minima, the universe inflates by a fraction of an $e$-fold.  After many
hundreds of tunneling events, the universe has inflated by the 60 (or so)
$e$-folds required to resolve the cosmological problems. At each stage, the phase
transition is rapid enough that bubbles of true vacuum intersect one another and percolation is complete.  Thus
the failure of "old" inflation \cite{Guth:1980zm}  to reheat is avoided, and "graceful exit" is achieved.

The authors of ref. \cite{Chialva:2008zw} took a different approach to chain inflation; they assumed that, during the course of evolution, the universe is separated
into patches, each in a different phase. These patches are distributed in a homogeneous manner. They are separated by domain walls whose collision will produce radiation.
Assuming that nucleation and thermalization occurs efficiently in each step, they studied the mechanism of production of density perturbations. They finally tried to realize chain inflation in the context of flux-compactified string theory, using the complex structure moduli as the agents. Their stringy models are motivated by their studies of string theory landscape \cite{Danielsson:2006xw,Chialva:2007sv}, where it was found out that related vacua in the context of flux compactification is a generic feature of landscape (see \cite{Ashoorioon:2008pj} for problems that may arise in this approach). Motivated by these studies, they considered two
landscape structures; in one the energy of the vacua changes linearly with the index of vacuum in the ladder, $\epsilon_i=i m_f^4$. In the second one, the energy is a quadratic function of the vacuum index, $\epsilon_i=\frac{i^2}{2}m_f^4$. The authors also obtained a formula for the amplitude of density perturbations which is different from the slow-roll one, by a factor of $\frac{1}{\sqrt{3}}$ in the denominator \footnote{Please note that we have used  $m_{\rm P}=(G)^{-\frac{1}{2}}$, which is different from the reduced Planck mass $M_{\rm P}=(8\pi G)^{-1/2}$ that was used in \cite{Chialva:2008zw}. }:
\begin{equation}\label{amp-dens}
\Delta_{\cal R}^2\equiv\frac{k^3 P_{\cal R}(k)}{2\pi^2}=\frac{H^2}{\pi m_{\rm P}^2\frac{\varepsilon}{\sqrt{3}}}.
\end{equation}

However in \cite{Chialva:2008zw}, the authors did not properly investigate the implications of WMAP observations for the fundamental parameters of these two models. The authors of \cite{Chialva:2008zw} had matched the models with the observation for an arbitrary mode, disregarding the moment it had exited the horizon. In the approach of \cite{Chialva:2008zw} to data-matching, the WMAP values of spectral index and density perturbation amplitude had been used as the input of the analysis and the free parameters of the model were tuned to obtain the observed values. Thus, they had not realized that their constructed models have predictions for the value of spectral index, irrespective of the value of $m_f$.  Here, we exactly determine the moment inflation ends, $t_e$, and match the amplitude of density perturbations, $60$ e-folds before that, $t_{60}$. Then, we calculate the spectral index for the mode that exit at $t_{60}$. This interestingly turns out to be independent of $m_f$. Thus, the value of scalar spectral index turns out to be the "prediction" of the model rather than some observational input for the analysis.  This approach  has already been used to extract the predictions of slow-roll models.

The running of the scalar spectral index turns out to be quite small in these two models. However the amount of gravity waves at large scales is substantial enough to be seen at the CMBPOL experiment \cite{Baumann:2008aq}, even though it will be smaller than what is expected in slow-roll models with linear and quadratic potentials respectively.

The outline of paper is as follows: first we will review the results of \cite{Chialva:2008zw} on how the  density perturbations and scalar spectral
index could be derived in chain inflation. Then we investigate the implications of WMAP observations for two vacua structure which are motivated by string theory. We also comment on how general the relation between the value of the scalar spectral index and vacua structure is.
Finally, we argue how one can possibly distinguish chain inflation models from slow-roll models. We conclude that large scale probes of the CMB could be used to confine the structure of string theory vacua, similar to how they have been used so far to confine the shape of the potential in slow-roll models.

\section{Review of the Previous Results}

In the approach of \cite{Chialva:2008zw}, during the course of inflation, our universe is divided into patches with different vacuum energies. Therefore, the main contributions to the energy density of the universe during chain inflation come from the cosmological constant within each region, $\rho^{\mathcal{V}}$, and also the energy stored in the bubble walls. All small patches are distributed in a homogeneous way. Collision of the bubble walls will transform this vacuum energy into radiation:
\begin{equation}\label{rho-tot}
\rho(t)=\rho^{\mathcal{V}}(t)+\rho^{W}(t).
\end{equation}
Denoting the fraction of volume occupied by the vacuum $i$ by $p_i(t)$ and the energy weighted fraction stored in uncollided bubble walls of the $i-1$-th phase by
$\mathcal{F}_{i,i-1}$, we have:
\begin{eqnarray}\label{}
\rho^{\mathcal{V}}(t)&=&\sum_{i=0} \epsilon_i p_i(t),\\
\rho^W (t)&=& \sum_{i=1}\Delta \epsilon_i\sum_{j=0}^{i-1}p_j(t)\mathcal{F}_{i,i-1}(t).
\end{eqnarray}
We also denote the nucleation rate per unit four volume by $\Gamma_i$. Assuming the total number of vacua to be $N+1$ and the nucleation rate per unit time,
$\tilde{\Gamma}_i(t)\equiv \Gamma_i V^{\rm physical}(t)=\tilde{\Gamma}$, to be constant over time and independent of $i$, we have to solve the following coupled
set of differential equations to find $p_m(t)$:
\begin{eqnarray}\label{coupled-probs}
  \dot{p}_N &=& -\tilde{\Gamma} p_N,\\ \nonumber
  \dot{p}_{N-1} &=& -\tilde{\Gamma} p_{N-1}+\tilde{\Gamma} p_N, \\ \nonumber
 &\cdots& \\
  \dot{p}_0 &=& \tilde{\Gamma} p_1, \nonumber
\end{eqnarray}
where
\begin{equation}\label{prob-sum}
\sum_i p_i(t)=1.
\end{equation}
The solution to \eqref{coupled-probs} for $p_i(t)$, the volume fraction in the $i$-th phase, is
\begin{equation}\label{pm}
p_i(t)=\frac{(\tg t)^{N-i}}{(N-i)!} e^{-\tg t}.
\end{equation}
The uncollided fraction of bubble walls between the $i$-th and $i-1$-th phases is calculated via the relation \cite{Turner:1992tz}
\begin{equation}\label{uncol-fraction}
{\cal F}_{i,i-1}(t)=\frac{\int_0^t dt' \Gamma_i V^{\rm physical}(t,t')p_i(t')p_{i,{\rm un}}(t,t')}{\int_0^t dt' \Gamma_i V^{\rm physical}(t,t')p_i(t')},
\end{equation}
which yields the following result when the decay rate per unit time $\tg$ is assumed to be constant
\begin{equation}\label{uncol-frac-G-const}
{\cal F}_{i,i-1}=\frac{p_{i-1}(t)(N-i)!}{(N-i)!-\Gamma(N-i+1,t)}.
\end{equation}
Above, $p_{i,{\rm un}}(t,t')$ is the probability that a wall generated at time $t'$ is still uncollided at time $t$, or in other words the probability that
a point outside the wall is in the phase $i$:
\begin{equation}\label{prob-uncollided-tg}
p_{i,{\rm un}}(t,t')=\exp\left( -\int_0^t \tg_i -\int_0^{t'} \tg_i\right).
\end{equation}
Motivated by some studies of stringy theory landscape, \cite{Chialva:2008zw} suggested two models where the energy density in each vacuum takes the following
functional forms in terms of the vacuum index
\begin{equation}\label{epsilon-n}
\epsilon_i=\left\{\begin{array}{cc}
                    m_f^4 i & {\rm case~I,} \\
                    \frac{m_f^4}{2} i^2 & {\rm case~II.}
                  \end{array}\right.
\end{equation}
The flux-compactified type IIB string theory is a natural framework to build such models. The quadratic behavior arises when the axiodilaton is stabilized independently of the complex structure moduli, while the linear one comes about when the axiodilaton is stabilized together with the structure moduli \cite{Chialva:2008xh, Chialva:2008zw}. The series of minima are generated with the help of monodromy transformations by revolving around a singular point in the moduli space of Calabi-Yau compactification. For these two models, \cite{Chialva:2008zw} calculated explicitly the total energy fraction in the bubble walls, $\rho^W(t)$, and the average of vacuum energy density inside each bubble, $\rho^{\cal V}(t)$, to find the total vacuum energy
\begin{equation}\label{rho-V-two-mdoels}
\rho(t)\sim \left\{\begin{array}{cc}
                    m_f^4 i(t) & {\rm case~I} \\
                    \frac{m_f^4}{2} i(t)^2 & {\rm case~II,}
                  \end{array}\right.
\end{equation}
where $i(t)\equiv N+1-\frac{t}{\tau}$. One can define the equivalent of the first slow-roll parameter $\varepsilon \equiv -\dot{H}/H^2 $ in these two cases, which turns out to be
\begin{equation}\label{slow-roll}
\varepsilon= \left\{\begin{array}{cc}
                    \frac{1}{2i(t)H\tau} & {\rm case~I,} \\
                    \frac{1}{i(t)H\tau} & {\rm case~II.}
                  \end{array}\right.
\end{equation}
The spectrum of density fluctuations turns out to be given by \eqref{amp-dens}. Also the spectral index as a function of $\varepsilon$ is given by \footnote{In the following, we have used the results of \cite{Chialva:2008xh} for the scalar spectral index, in which the effect of the term $-\frac{1}{3}H^2$ in the equation of motion for scalar perturbations was taken into account.}
\begin{equation}\label{ns}
n_s\equiv 1+\frac{d \ln \Delta_{\cal R}^2}{d\ln k}=\left\{\begin{array}{cc}
                    1-\frac{17}{3}\varepsilon & {\rm case~I} \\ \\
                    1-\frac{14}{3}\varepsilon & {\rm case~II}
                  \end{array}\right.
\end{equation}\
Here, we will explicitly investigate the observational implications of these two models and will see that both of them have predictions for the scalar spectral index
which are independent of $m_f$ and are compatible with the WMAP results.

\section{Matching the Models with Observation}

The modes that are observed at the CMB scales left the horizon approximately  $60$ e-folds before the end of inflation. For these modes, the latest WMAP experiment \cite{Komatsu:2010fb} has measured the
amplitude of scalar perturbations and its corresponding spectral index to be $\Delta_{\cal R}^2= (2.441\pm 0.096) \times 10^{-9}$ and $n_s=0.963\pm 0.012$, respectively. Here
we determine the moment that inflation ends explicitly and will match the free parameters of the model with observation in order to find the correct amplitude for density
perturbations. We focus initially on the case where the energy density in each stage of the ladder is a linear function of $i$.

\subsection{$\mathbf{\epsilon_{\it i}=\textit{i}~m_f^4}$}

As in the slow-roll case, the end of inflation is designated by $\vepsn=1$. At this moment the acceleration of the space-time becomes slow enough to let the bubble walls collide, coalesce and transform into radiation. Using eq.\eqref{rho-V-two-mdoels}, we can determine the Hubble parameter as a function of time via the Friedmann equation:
\begin{equation}\label{H-t}
H(t)=\frac{m_f^2}{m_{\rm P}}{\left(\frac{8\pi}{3}\right)}^{1/2}\sqrt{N+1-\frac{t}{\tau}}
\end{equation}
The end of inflation is set by equation $\vepsn=1$, which determines $t_f$ to be
\begin{equation}\label{tf}
t_f=(N+1)\tau-{\left(\frac{3 \tau m_{\rm P}^2}{32 \pi m_f^4}\right)}^{1/3}.
\end{equation}
As stated above, the observed CMB scales exit the horizon around $60$ e-folds before the end of inflation. For the above $t_f$, number of e-foldings as a function of the initial time
$t_i$ is
\begin{eqnarray}\label{Ne-t}
N_e(t_i,t_f)&=&\int_{t_i}^{t_f} H~dt\\\nonumber
             &=&\frac{4\sqrt{6\pi}}{9} \frac{m_f^2}{m_{\rm P}} \sqrt{N+1-\frac{t_i}{\tau}}\left[(N+1)\tau-t_i\right]
             -\frac{1}{3}.
\end{eqnarray}
$60$ e-folds before the end of inflation is determined by equation $N_e(t_{60},t_f)=60$, which yields:
\begin{equation}\label{t60}
t_{60}=(N+1)\tau-{\left(\frac{98283~\tau m_{\rm P}^2}{32\pi m_f^4}\right)}^{1/3}.
\end{equation}
Requiring $t_{60}\geq 0$, one obtains the minimum number of vacua needed to obtain $60$ e-foldings in this specific model of chain inflation:
\begin{equation}\label{Nmin-tau}
N_{\rm min}= {\left(\frac{98283 m_{\rm P}^2}{32\pi m_f^4 \tau^2}\right)}^{1/3}-1
\end{equation}
We now require that $\Delta_{\cal R}^2(t_{60})$ matches the central value of the WMAP 7 year results, $2.441\times 10^{-9}$ \cite{Komatsu:2010fb}
. This establishes a relation
between the energy gap of the adjacent vacua, $m_f^4$, and the tunneling time, $\tau$
\begin{equation}\label{tau-mf}
\tau\simeq 6.2669\times 10^{18} \frac{m_f^4}{m_{\rm P}^5}.
\end{equation}
Plugging this result into \eqref{Nmin-tau} yields the minimum number of vacua needed to realize $60$ e-folds of inflation as a function of the energy gap between the vacua:
\begin{equation}\label{Nmin-mf}
N_{\rm min}=\frac{2.9198\times 10^{-12} m_{\rm P}^4}{m_f^4}-1.
\end{equation}
This relation indicates that as we decrease the energy gap between the adjacent vacua, more and more minima are required to obtain enough inflation. From the fact that $N_{\rm min}(m_f)$ has to be bigger than or equal to one, it is obtained that
\begin{equation}\label{mfmax}
m_f\leq 1.1\times 10^{-3}~m_{\rm P}
\end{equation}
For such an energy gap, the nucleation time is $9.1491 \times 10^6 m_{\rm P}^{-1}$. Of course, as we will see later, the requirement of thermalization at each step will
lead to a more stringent constraint on the energy gap. Such a constraint will lead to much larger lower bound on the number of vacua, which justifies the statistical
approach of \cite{Chialva:2008zw} who had assumed that the number of minima was large.

The scalar spectral index, $n_s(t_{60})$, at such scales could be found through the relation \eqref{ns} using eq.\eqref{t60}. Interestingly, we find that its value is
independent of $m_f$ and $\tau$:
\begin{equation}\label{ns-nsquared}
n_s(t_{60})=0.9687
\end{equation}
which is within the $2\sigma$ level of WMAP seven year results. This is different but very close to the prediction of a linear potential, $V(\phi)=\kappa\phi$, for the scalar spectral index.
For such a linear potential, the scalar spectral index at large scales is $0.975103$. This difference is due to interaction of vacuum and radiation components of energy density, which is the characteristic of chain inflation. In principle a very precise measurement of the CMB at large scales should be able to
differentiate between the slow-roll linear potential and its chain inflation counterpart, $i~m_f^4 $.

The overall shape of the potential in the chain inflation model could be far from shallowly linear. This is in contrast with the slow-roll model which should be quite shallow to sustain inflation and produce the right amount of perturbations, $\kappa\simeq 1.69\times 10^{-13} M_{\rm P}^3$. The only requirements in the chain inflation model, are having a sufficient number of minima to get $60$ e-folds of inflation, \textit{i.e.} constraint \eqref{Nmin-mf}, and having the right nucleation rate at each step, \textit{i.e.} equation \eqref{tau-mf}. The second requirement could be achieved even with a potential whose overall shape is far from flat. This is because the nucleation rate depends on the energy gap between the minima, the separation of the minima in the field space and the height of the barrier that separates the metastable minimum from the true one. Fixing the first parameter in our problem, we still have two other parameters that we can play with to achieve the right nucleation rate. Therefore the macroscopic shape of the potential could be far from linear shape. In fact, it could be such that it would not be able to sustain inflation in the slow-roll sense at all.

Requiring thermalization at each step reveals a stronger constraint on the energy gap between the vacua. In \cite{Chialva:2008zw}, it was understood that in order to get
percolation and large scale thermalization at each step of chain inflation, one has to assume that $\tg>3H$. If one assumes that chain inflation has started from the
Planckian energies, $m_{\rm P}^4$, this will lead to the following upper bound on the mass gap between the consecutive vacua:
\begin{equation}\label{}
m_f<2.6069\times 10^{-5}~m_{\rm P}.
\end{equation}
On the other hand, if one assumes that the number of vacua has been enough to get $60$ e-folds, this constraint will be weakened as
\begin{equation}\label{}
m_f<3.2203\times 10^{-4}~m_{\rm P}.
\end{equation}
One should note that for the above energy gap, the required number of minima to get a sufficient number of e-foldings in chain inflation is $\simeq 270$ which is large enough to justify the statistical
approach of \cite{Chialva:2008zw}. Of course the existing number of vacua could be large enough to tile the energy difference between zero and the Planckian energy density, {\it i.e.} $N=\left(\frac{m_{\rm P}}{m_f}\right)^{4}+1$. Then one would obtain larger amount of inflation than what is needed to solve the problems of the Standard Big Bang Cosmology.
\begin{figure}
  \includegraphics[width=80mm,height=70mm]{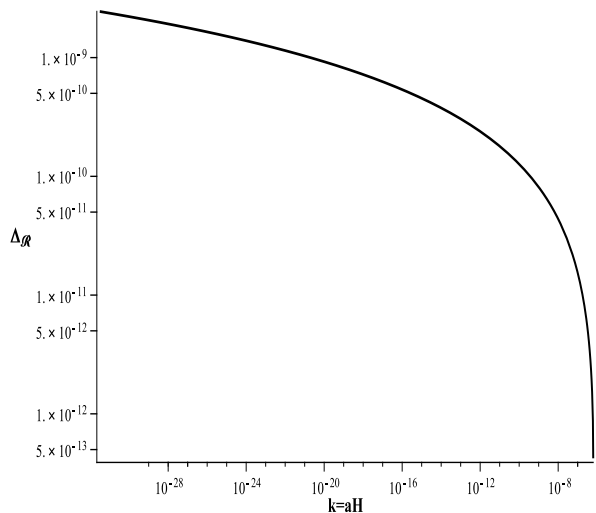}\\
  \caption{The graph shows the scale dependence of the scalar power spectrum for a landscape with $\epsilon_i=i m_f^4$.
  The scalar spectral index turn out to be $0.972375$. }\label{density-mf=10^-5}
\end{figure}

\begin{figure}
  \includegraphics[width=80mm,height=70mm]{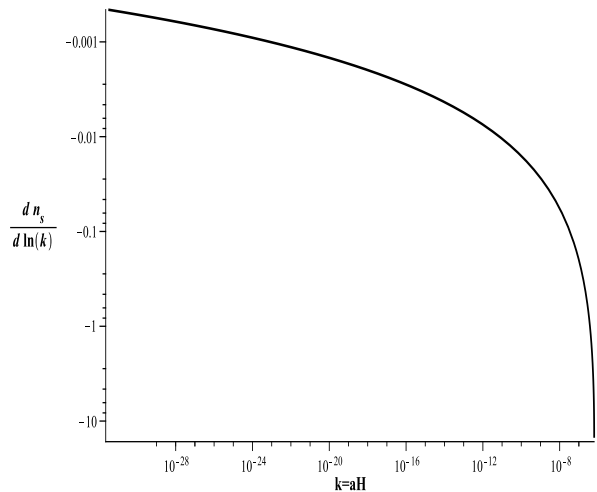}\\
  \caption{The graph shows the scale dependence of running of scalar spectral index. Even within the range of scales that will be accessible to small scale probes, the running is small. }\label{running-ns-fig}
\end{figure}

We have plotted the primordial scalar spectrum for the modes that exit the horizon during the last $60$ e-folds of inflation for $m_f=10^{-5}~m_{\rm P}$ in
{\bf Fig.}(\ref{density-mf=10^-5}). Even though the final
spectrum is independent of the number of vacua as long as $N\geq N_{\rm min}(10^{-5}m_{\rm P})$, we have assumed that we have a sufficient number of minima to fill
the energy interval from zero to Planckian energies, \textit{i.e.} $N\simeq 10^{20}$. The spectrum is almost scale-invariant at large scales but it
has a very red spectrum at small scales. We have also plotted the running of the spectral index, $\frac{d n_s}{d\ln k}$ vs. $k$,
for the modes that exit within the last $60$ e-folds, see {\bf Fig.}(\ref{running-ns-fig}). In terms of the Hubble parameter and $\varepsilon$, the running is:
\begin{equation}\label{running}
\alpha\equiv\frac{dn_s}{d\ln k}=-\frac{5\dot{\varepsilon}}{H}.
\end{equation}
The running is negative and negligible for our Hubble
size modes. In fact for $m_f=10^{-5} m_{\rm P}$, $\frac{dn_s}{d\ln k}=-4.5768\times 10^{-4}$ at $k_{\star}=0.002~ {\rm Mpc^{-1}}$ which is consistent with WMAP 5 years results.
On the other hand, for the modes that exit in the last few e-folds of inflation and are important for the black hole formation, the running is very negative, $\simeq -15$.
One might hope that in the observable intermediate scales the running would become substantial. However, even for the modes that exit $48$ e-folds before the end of inflation and
contribute to Lyman-$\alpha$ forest, {\it i.e.} $k\simeq 500~{\rm Mpc^{-1}}$, $\frac{dn_s}{d\ln k}=-7.1343\times 10^{-4}$, which is below the detectability limit.

Since the minimum number of minima which is required to achieve enough inflation increases like \eqref{Nmin-mf} as $m_f$ decreases, the energy scale at which inflation happens does not
decrease by decreasing $m_f$. In fact, it turns out that for the nucleation time as given in \eqref{tau-mf}, the energy scale of inflation at $t_{60}$ is independent of $m_f$ and is close to the GUT scale:
\begin{equation}\label{}
H(t_{60})=4.9458\times 10^{-6}~m_{\rm P}
\end{equation}
This corresponds to $r\simeq 0.05103$ for gravity waves at the horizon scale, which is certainly observable by probes like the CMBPOl \cite{Baumann:2008aq}. This is smaller than the tensor/scalar ratio one expects from a linear potential, $V(\phi)=\kappa\phi$, by a factor of $\frac{1}{\sqrt{3}}$. As we will explain extensively later this suppression, through the modification it induces in the form of consistency relation, can be used to distinguish chain inflation from its slow-roll counterparts. One should note that aside from quantum fluctuations of the metric, in chain inflation, we also have gravity waves from bubble collision. Another paper has already been dedicated to study that effect and I invite the enthusiastic reader to study it \cite{Ashoorioon:2008nh}.

\subsection{$\mathbf{\epsilon_{\it i}=\frac{{\it i}^2}{2} m_f^4}$}

The analysis of a quadratic structure in the vacua goes hand in hand with the linear case. Therefore we will skip some of the details.
In this case the Hubble parameter decreases linearly as a function of cosmic time
\begin{equation}\label{H-n2}
H(t)=\frac{m_f^2}{m_{\rm P}}{\left(\frac{4\pi}{3}\right)}^{1/2}\left(N+1-\frac{t}{\tau}\right).
\end{equation}
We have to impose the density perturbations amplitude $60$ e-folds before the end of inflation,
which is determined by the condition, $\varepsilon=1$. This will determine the nucleation rate, $\tau$, in terms of $m_f$
\begin{equation}\label{tau-mf-n2}
\tau\simeq 6.7679\times 10^{12} \frac{m_f^2}{m_{\rm P}^3}.
\end{equation}
The scalar spectral index at the current Hubble scale $k_{\ast}=0.002~{\rm Mpc^{-1}}$ can
be found using the second equation in \eqref{ns}. Interestingly, the result again turns out to be
independent of $m_f$:
\begin{equation}\label{ns-n2}
n_s=0.9614.
\end{equation}
This is also close but different from what is obtained for a slow-roll model with potential $\frac{m^2}{2}\phi^2$.

Demanding to obtain at least $60$ e-folds of inflation in this model, leads to the following lower bound on the number of minima as a function of $m_f$
\begin{equation}\label{Mmin-mf-n2}
N_{\rm min}=\frac{2.9556\times 10^{-6}m_{\rm P}^2}{m_f^2}-1.
\end{equation}
If one assumes that thermalization has occurred after each phase transition and inflation has started at Planckian energies, one would obtain the following upper bound on
$m_f$
\begin{equation}\label{mf-upper-bound}
m_f\leq 6.5396\times 10^{-7}~m_{\rm P}.
\end{equation}
Lowering our ambition to have only $60$ e-folds, increases the upper bound on $m_f$ to $9\times 10^{-5}~m_{\rm P}$.
The values of the Hubble parameter at the beginning and at the end of inflation turn out to be $6.049\times 10^{-6}~m_{\rm P}$ and $5.4991\times 10^{-7}~m_{\rm P}$, respectively, which are independent of $m_f$ and $N$. These values for the Hubble parameter correspond to $r\simeq 0.0763$ at horizon scales, which is again smaller than its slow-roll counterpart, $\frac{1}{2}m^2\phi^2$, by a factor of $\frac{1}{\sqrt{3}}$. Nonetheless, the gravity wave signal should be observable at CMBPOL.
Running, $\frac{d n_s}{d\ln k}$ at $k=k_{\ast}$ is larger than the model with linear structure, $-6.83\times 10^{-4}$, but is still too small. At the scales accessible
to Lyman-$\alpha$ forest the running is $-1.0628\times 10^{-3}$, which is still below the detectability limit.

One may wonder how general the relation between the value of the scalar spectral index and the vacuum structure is. In order to answer this question, we consider the vacua structure with general power-law dependence, $\epsilon_i=\frac{m_f^4}{c!}i^c$, even though the stringy realization of these vacua structures for $c>2$ is still lacking. These are structures that had already been discussed in \cite{Chialva:2008xh}. For these vacua structures, the dependence of $n_s$ on the slow-roll parameter is:
\begin{equation}\label{ns-epsilon-different-c}
n_s=1-\left(\frac{11}{3}+\frac{2}{c}\right)\varepsilon(t)
\end{equation}
Following the same procedure for data-matching, the scalar spectral index turns out to be independent of $m_f$. Its value depends on the parameter $c$ as follows:
\begin{equation}\label{ns-c}
n_s=\frac{2(177+86 c)}{3(120+61c)}
\end{equation}
WMAP seven year results for the scalar spectral index shows that $c>5$ is ruled out with $68\%$ C.L.

\section{Distinguishing Chain Inflation from its Slow-roll Counterparts}

We noticed that the scalar spectral index in the chain inflation models is in general different from its slow-roll counterparts. In fact this point was first noticed in \cite{Chialva:2008xh}. This is due to the interaction between radiation and vacuum components of energy density. Another criterion which was discussed in \cite{Chialva:2008xh} to be the signature of chain inflation is the super-imposed oscillations that are due to break-down of the coarse-graining approach on scales comparable to the size of the nucleating bubble, $r_b$. The size of effect was estimated to be proportional to $r_b H$, where $H$ is the Hubble parameter during the inflation.

However a more promising criterion for distinguishing chain inflation from the usual chaotic slow-roll inflation lies in the specific form of the density perturbations formula that has an extra factor of $\frac{1}{\sqrt{3}}$ in the denominator, see eq. \eqref{amp-dens}. This leads to a modification of consistency relation between the tensor and scalar spectra produced from chain inflation:
\begin{equation}\label{chain-consistency}
r=-\frac{8 n_T}{\sqrt{3}},
\end{equation}
where $n_T$ is the tensor spectral index. The extra factor of $\sqrt{3}$ is indication of the fact that the role of kinetic energy, whose sound speed is one, is played by radiation which has $c_s=\frac{1}{\sqrt{3}}$. The consistency relation between tensor and scalar spectra ,
\begin{equation}\label{consistency}
r=-8 n_T,
\end{equation}
is the firm prediction of single-field slow-roll models and is reflective of the fact that these two spectra are derived from the same inflationary potential. The relation was first noticed by \cite{Liddle:1992wi} in the context of power-law spectra and later found to be true in more general situation \cite{Copeland:1993jj}. As discussed previously, the extra factor of $\sqrt{3}$ in the denominator of the R.H.S. of \eqref{chain-consistency} corresponds to fewer gravity waves at large scales \footnote{Trans-Planckian effects can also cause the violation of consistency relation \cite{Ashoorioon:2005ep} which cast an oscillatory term in the R.H.S. of \eqref{consistency}.}. This smaller amount of gravity waves at large scales, in principle, could make the detection of its scale dependence harder. However for a given value of $r$ bigger than $0.01$, which is detectable by balloon-borne experiments, the corresponding value of tensor spectral index is larger by a factor of $\sqrt{3}$ for chain inflation models. For example for $r=0.05$, which is the prediction of chain inflation model with linear structure, the predicted value of $n_T$ is around $0.01$, which is larger than that of single-field slow-roll model with the same value of $r$, {\it i.e.} $n_T=\frac{0.05}{8}=0.0625$. In this way, the modified consistency relation facilitates the detection of scale-dependence of gravity waves.

Lensing of the $E$ mode by density perturbations along lines of sight from the last scattering surface limits the detectibility of $B$ modes at $r_{\rm lim}=2.6\times 10^{-4}$ \cite{Knox:2002pe}. As mentioned above, for $r\simeq \mathrm{few} \times 0.01$, the measurement of tensor spectral index should be very difficult if the single-field consistency relation holds. Nonetheless, it is shown that large departures from the consistency relation can be seen if $r\gtrsim 10^{-3}$ \cite{Song:2003ca}. In particular, \cite{Song:2003ca} shows that observing any nonzero value for $n_T+r/8$ in the region $r<0.166$ \footnote{Please note that my normalization of $r$ is different from theirs by a factor of $8/4.8$.}, can confidently lead us to exclude the single-field slow-roll model, as the loop correction (such as the one discussed in \cite{Kaloper:2002cs}) to the single-field consistency relation is quite small in this interval of $r$. All chain inflation models predict the violation of the consistency relation with $r$ bigger than $0.05$, but smaller than $0.166$. Therefore distinguishing chain inflation from slow-roll models should be observationally possible.

Knowing that the realization of slow-roll inflation within string theory is quite a difficult task \cite{McAllister:2005mq} (see \cite{Ashoorioon:2009wa,Silverstein:2008sg} for some attempts) and noting that metastable vacua are abundant within the string theory landscape, chain inflation might seem a promising alternative approach to realize inflation within string theory. In this paper, we focused on a class of flux-compactified theories in which a series of  minima are generated by monodromy transformations. In particular we focused on two models  in which linear and quadratic vacua structure arises. We demonstrated that these two models have specific predictions for the value of the scalar spectral index and could be distinguished from their slow-roll counterparts by the violation of the consistency relation between tensor and scalar spectra.

Seventeen years ago, it was argued that with the knowledge of tensor and scalar perturbations, one is in principle able to reconstruct the inflationary potential \cite{Copeland:1993jj,Copeland:1993ie}. Some work was done on the reconstruction of inflaton potential based on this observation. The tensor spectrum plays a crucial role in such reconstructions. Since chain inflation with different vacua structure has specific predictions for the tensor and scalar spectra, one may similarly hope that probing the scalar and tensor perturbations at large scales can help us to confine the global structure of stringy vacua. Noting that future CMB experiments are able to determine $n_s$ within $\sigma(n_s)=0.0024$ \cite{Kaplinghat:2003bh} and extracting experimental information from the string theory landscape is a formidable task, this is a promising prospect.

\section*{Acknowledgments}
The author is thankful to U. Danielsson, D. Chialva and K. Freese for useful discussions and to Susha Parameswaran for revising and editing the manuscript. This work is supported by the G\"{o}ran Gustafsson Foundation.

\end{document}